\documentclass[journal,10pt]{IEEEtran}
\usepackage{mathrsfs}
\usepackage{amsfonts}
\usepackage{graphicx,cite,epsfig,amssymb,amsmath}
\usepackage{color,xcolor}
\definecolor{red}{rgb}{1.00, 0.00, 0.00}  
\usepackage{pifont}
\usepackage{stmaryrd}
\usepackage{setspace}
\usepackage{subfigure}
\usepackage{cite}
\usepackage{array}
\usepackage{float}
\usepackage{multirow}
\usepackage{algorithm,algpseudocode}
\usepackage{epstopdf}
\usepackage{mathtools}
\usepackage{diagbox}
\usepackage{booktabs}
\usepackage{verbatim}

\providecommand{\algorithmname}{Algorithm}
\floatname{algorithm}{\protect\algorithmname}
\newcommand{\bm}[1]{\mbox{\boldmath{$#1$}}}
\newcommand{\tabincell}[2]{\begin{tabular}{@{}#1@{}}#2\end{tabular}}

\usepackage{arydshln}
\usepackage{enumitem}

\begin{document}
	
\title{Accelerating Generalized Benders Decomposition for Wireless Resource Allocation}
\author{Mengyuan Lee, Ning Ma, Guanding Yu, and Huaiyu Dai
	\thanks{M. Lee  and  G. Yu are with College of Information Science and Electronic Engineering, Zhejiang University, Hangzhou 310027, China. e-mail: \{mengyuan\_lee,  yuguanding\}@zju.edu.cn. (\emph{Corresponding author: Guanding Yu})}
	\thanks{N. Ma is with the College of Computer Science and Technology, Zhejiang University, Hangzhou 310027, China. e-mail: 3170101236@zju.edu.cn.}
	\thanks{H. Dai is with the Department of Electrical and Computer Engineering, North Carolina State University,
		Raleigh, NC 27606 USA. e-mail: hdai@ncsu.edu.}
	\thanks{This work was done while M. Lee was a visiting PhD student at NC State University.}
	}
\maketitle

\begin{abstract}
Generalized Benders decomposition (GBD) is a globally optimal algorithm for mixed integer nonlinear programming (MINLP) problems, which are NP-hard and can be widely found in the area of wireless resource allocation. The main idea of GBD is decomposing an MINLP problem into a primal problem and a master problem, which are iteratively solved until their solutions converge. However, a direct implementation of GBD is time- and memory-consuming. The main bottleneck is the high complexity of the master problem, which increases over the iterations. Therefore, we propose to leverage machine learning (ML) techniques to accelerate GBD aiming at decreasing the complexity of the master problem. Specifically, we utilize two different ML techniques, classification and regression, to deal with this acceleration task. In this way, a cut classifier and a cut regressor are learned, respectively, to distinguish between useful and useless cuts. Only useful cuts are added to the master problem and thus the complexity of the master problem is reduced. By using a resource allocation problem in device-to-device communication networks as an example, we validate that the proposed method can reduce the computational complexity of GBD without loss of optimality and has good generalization ability. The proposed method is applicable for solving various MINLP problems in wireless networks since the designs are invariant for different problems.
\end{abstract}

\begin{IEEEkeywords}
Machine learning, generalized Benders decomposition,  device-to-device communications, resource allocation, mixed integer nonlinear programming
\end{IEEEkeywords}

\section{Introduction}
Resource allocation is a popular and important topic in wireless networks \cite{Han}. However, wireless resource allocation problems generally involve two kinds of variables, i.e., the continuous and discrete ones. Therefore, most of them are formulated as mixed integer nonlinear programming (MINLP) problems, which can be widely found in various existing works \cite{chen,fplinq,MECO,UAV,relay}. 

Unfortunately, MINLP problems are generally NP-hard and no efficient globally optimal algorithm exists yet. Available globally optimal algorithms for MINLP problems, such as the branch and bound (B\&B) algorithm \cite{bb} and the generalized Benders decomposition (GBD) \cite{GBD}, suffer from the exponential worst-case computational complexity. Therefore, sub-optimal and heuristic algorithms of low complexity are widely used for resource allocation in wireless networks based on various mathematical techniques, such as game theory \cite{gametheory}, graph theory \cite{graphtheory}, greedy search \cite{greedysearch}, and some other iterative methods \cite{iterative}. However, their performance has no guarantee because the gaps between the optimal solutions and the sub-optimal ones are difficult to control.

Inspired by the success of machine learning (ML) in many fields, the wireless communications community has recently turned to ML to obtain more efficient methods for resource allocation problems, such as power allocation \cite{shi,mlop2,mlop3,lorm,learntobranch,deepfolding}, link scheduling \cite{spatiallearning,graphnn}, and user association\cite{user1}. All aforementioned works can be classified into three different learning paradigms. The first one is the end-to-end learning paradigm. For this learning paradigm, the input/output relation of a given resource optimization problem is regarded as a “black box”, which is directly learned by ML techniques especially the deep neural networks (DNNs). However, this learning paradigm is effective for resource allocation with only one kind of output variables \cite{shi,deepfolding,spatiallearning,graphnn} and does not work well for MINLP problems. The second one is the reinforcement learning paradigm. Because labeled samples that are difficult to obtain in wireless networks are not needed in this learning paradigm, it is widely used in many wireless resource allocation scenarios \cite{mlop2,mlop3,user1}. However, problem-dependent designs are needed and how to choose appropriate designs is tricky, which needs iterative processes of trial and error. The third one is the algorithm specific learning paradigm, which aims at accelerating some existing algorithms for optimization problems. Specifically, it uses ML techniques to exploit the specific algorithm structures. For example, authors in \cite{lorm,learntobranch} have utilized imitation learning to learn the best prune policy to accelerate the B\&B algorithm.  It can achieve satisfactory performance even without problem-dependent designs. However, the accelerated B\&B algorithm in \cite{lorm,learntobranch} suffers from loss of optimality.

To circumvent problem-dependent design and avoid loss of optimality, we propose to use ML techniques to accelerate the GBD algorithm in this paper. GBD algorithm is widely used for resource allocation in wireless networks \cite{r2_add1,r2_add2,r2_add3,r2_add4}. It performs equally well as the B\&B algorithm but has a simpler algorithm structure.  The main idea of GBD is decomposing an MINLP problem into a primal problem and a master problem. These two problems are iteratively solved until their solutions converge.  At each iteration,  the primal problem is first solved, and a cut, i.e., a new constraint, is generated according to its result. Then the cut is added to the master problem. Finally, the master problem is solved. Over the iterations, the number of added cuts increases and the  master problem becomes more  and more time-consuming. Therefore, the high complexity of the master problem is the main bottleneck of GBD. To accelerate GBD, many mathematical methods have been proposed. One of the widely used methods is the multi-cut GBD \cite{multicut}, where multiple cuts instead of one  cut are generated at each iteration to reduce the number of required iterations. In this way, fewer master problems are needed to be solved and the total computational time can be reduced. However, not all the cuts generated during multi-cut GBD contribute equally to the convergence. Some cuts would not bring much change to the convergence but increase the complexity of master problem. Therefore, we propose to use ML techniques to distinguish between useful and useless cuts, and only add useful cuts to the master problem. In this way, the complexity of master problems increases in a slower speed and GBD can be further accelerated. Specifically, we utilize two different ML techniques, classification and regression, to deal with this acceleration task. Then we design five features to characterize the generated cuts. These features are invariant  and applicable for different resource allocation problems. We also propose different data collection methods, label definition methods, as well as training and testing processes for these two kinds of ML techniques. In this way, a cut classifier and a cut regressor can be  learned in the supervised manner, respectively. Finally, we use a joint subcarrier and power allocation problem in  device-to-device (D2D) communication networks  as an example to verify the effectiveness and generalization ability of the proposed method. 

Actually, cut deletion or clean-up strategies are important when multiple cuts are added to the master problems at each iteration \cite{bdsurvey}. Most existing works use mathematical methods to develop cut clean-up strategies. A similar work that has leveraged ML techniques to develop cut clean-up strategies can be found in \cite{svmgbd}, where the authors have accelerated Benders decomposition \cite{BD} especially for the two-stage stochastic programming problems. Specifically, the authors have used ML techniques to evaluate cuts and their proposed method is instance-specific, which means a specific model is needed for each instance and the training process is conducted online. However, reusing models for different problems and offline training are generally preferred in wireless networks.  Moreover, the two-stage stochastic programming problems are very different from the MINLP problems and the cuts generated in GBD are more difficult to be characterized than those in Benders decomposition. Therefore, the method in \cite{svmgbd} may not be directly applied for wireless resource allocation. 

In brief, our main contributions are summarized as follows.

\begin{itemize}[leftmargin=3mm]
	\item We incorporate ML techniques to accelerate GBD. The proposed method is a practically efficient and globally optimal algorithm. It is applicable for various MINLP resource allocation problems in wireless networks, thus avoiding tricky problem-dependent designs in individual cases. Extensive simulation results suggest that the proposed method can reduce the time complexity of GBD without loss of optimality and has strong generalization ability.
	\item As a universal definition for useful cuts is not available \cite{nodefine}, the obtained binary labels for generated cuts are inevitably noisy, which has a significant impact on the performance. To alleviate the impact of noise, we propose to utilize two different ML techniques, i.e, classification and regression, to develop cut clean-up strategies and accelerate the GBD algorithm. Moreover, simulation is conducted to analyze and compare the advantages and disadvantages of these two methods.
	\item To solve these two ML problems, we carefully design five problem-independent features to describe the cuts generated in GBD, together with the corresponding data collection methods, label definition methods, as well as training and testing processes.
	\item Moreover, based on theoretical analysis on noisy labels in binary classification, we provide some interesting insights for choosing hyperparameters even without knowing the accurate noise in the training samples, thus avoiding the time-consuming process of trial and error in the traditional hyperparameter selection procedure.
\end{itemize}

The rest of this paper is organized as follows. In Section II, we introduce the general mathematical formulation for wireless resource allocation and give a brief introduction to GBD. In Section III, we use ML techniques to accelerate GBD. The testing results of the proposed method are presented in  Section IV. Finally, we conclude this paper in Section V.

\section{GBD for Wireless Resource Allocation }
In this section, we first introduce the general mathematical formulation for wireless resource allocation problem. Then, we introduce the basic idea of the classic GBD algorithm.  Finally, we analyze the drawbacks of GBD and introduce one improved variant, i.e., the multi-cut GBD.

\subsection{Problem Formulation for Wireless Resource Allocation}
Resource allocation in wireless networks generally involves two kinds of variables: the continuous ones such as data rate and transmit power, and the discrete ones such as user selection and subcarrier assignment. Thus, many wireless resource allocation problems can be formulated as the following generic form of mixed integer optimization\footnote{Maximization problems can be transformed to minimization problems in a straightforward way.}
\begin{eqnarray*}\label{Mresource}
	\hspace{8.8em}  \min_{\{\bm{x}, \bm{y}\}} f(\bm{x}, \bm{y}),\hspace{9em}
	\eqref{Mresource_obj} \label{Mresource_obj}
\end{eqnarray*}
\begin{subequations}
	subject to
	\vspace{-1em}
	\begin{align}
	\bm{G}(\bm{x}, \bm{y}) \leq 0,\label{resource_sub1}
	\end{align}
	\vspace{-2em}
	\begin{align}
	\bm{x} \in \mathcal{X},\label{resource_sub2}
	\end{align}
	\vspace{-2em}
	\begin{align}
	\bm{y} \in \mathcal{Y},\label{resource_sub3}
	\end{align}
\end{subequations}
where $f(\cdot,\cdot)$ is the objective function such as energy consumption and communication delay, $\bm{x}$ is the vector of $n_1$ continuous variables, $\bm{y}$ is the vector of $n_2$ discrete variables, and $\bm{G}(\cdot,\cdot)$ represents the vector of constraints defined on $\mathcal{X} \times \mathcal{Y} \subseteq \mathbb{R}^{n_1} \times  \mathbb{Z}^{n_2}$, such as power constraints and data rate requirements. 

Problem (\ref{Mresource}) can be widely found in the study of wireless networks, such as joint user association and power allocation in carrier aggregation systems \cite{chen},  optimal mode selection in D2D networks \cite{fplinq},  resource allocation in multi-user mobile-edge computation offloading \cite{MECO}, trajectory planning and recourse allocation for unmanned aerial vehicles \cite{UAV},  as well as joint power allocation and relay selection in amplify-and-forward cooperative communication system \cite{relay}. 

Generally, Problem (\ref{Mresource}) is NP hard and it is computationally challenging to obtain its optimal solution.  According to the linearity of $f(\cdot,\cdot)$ and  $\bm{G}(\cdot,\cdot)$ with fixing $\bm{y}$, Problem (\ref{Mresource}) can be further categorized into two kinds: mixed integer linear programming (MILP) problems and MINLP problems. For MILP problems, both $f(\cdot,\cdot)$ and  $\bm{G}(\cdot,\cdot)$ with fixing $\bm{y}$ are supposed to be linear functions. Unfortunately, this condition is not met by most wireless resource allocation problems. Therefore, MILP problems are less often encountered than MINLP problems in wireless networks. The optimal algorithms for solving the MINLP problems include the B\&B algorithm and the GBD algorithm. Generally, GBD has a simpler algorithm structure than the B\&B algorithm. In the following, we introduce the basic idea of the GBD algorithm for MINLP problems.

\subsection{Brief Introduction of GBD}
The main idea of GBD is decomposing an MINLP problem into two sub–problems: a primal problem and a master problem.  These two problems are iteratively solved until their solutions converge. To guarantee the convergence and the global optimality of the GBD algorithm, the MINLP problem in (\ref{Mresource}) should satisfy two conditions \cite{GBD}:
\begin{itemize}[leftmargin=4mm]
	\item Convexity:  the problem should be convex on $\bm{x}$ given the discrete variables $\bm{y}$;
	\item Linear separability: the problem should be linear on $\bm{y}$ given the continuous variables $\bm{x}$. 
\end{itemize}
Otherwise, GBD algorithm may fail to converge or converge to a local optimum. Note that linear separability is satisfied by many wireless resource allocation problems, such as Problem (\ref{Md2d}) in Section IV.A. In this paper, we only focus on the MINLP problems that satisfy the above two conditions, for which the GBD algorithm always convergences to the global optimum.  As for the non-convex MINLP problems, many variations of GBD have been proposed \cite{nonconvex1, nonconvex2,nonconvex3}. How to accelerate those modified GBD algorithms is interesting for future work.

For a convex and linear separable MINLP problem, the primal problem is also convex and can be solved by various convex optimization techniques. Specifically, the primal problem at the $i$-th iteration finds the optimal $\bm{x}$ by fixing the discrete variables $\bm{y}$ on a specific value $\bm{y}^{(i-1)}$ (or a given initial $\bm{y}^{(0)}$) and can be written as
\begin{eqnarray*}\label{Mprimal}
	\hspace{8em}\min_{\bm{x} \in \mathcal{X}} f(\bm{x}, \bm{y}^{(i-1)}),\hspace{8.3em}
	\eqref{Mprimal_obj} \label{Mprimal_obj}
\end{eqnarray*}
\begin{subequations}
	subject to
	\vspace{-1em}
	\begin{align}
	\bm{G}(\bm{x}, \bm{y}^{(i-1)}) \leq 0.\label{primal_sub1}
	\end{align}
\end{subequations}
Note that some specific value $\bm{y}^{(i-1)}$ may lead to an infeasible primal problem (\ref{Mprimal}). If Problem (\ref{Mprimal}) is feasible, the corresponding solution is denoted as $\bm{x}^{(i)}$ and we add the iteration index $i$ into the feasible primal problem index set, $\mathcal{F}$, as a new element. Clearly, the optimal value $f(\bm{x}^{(i)},\bm{y}^{(i-1)})$
provides an upper bound to the original Problem (\ref{Mresource}). Given that Problem (\ref{Mprimal}) is convex, solving the problem also provides the optimal Lagrange multipliers vector, i.e., the optimal dual variables vector, $\bm{\mu}^{(i)}$, for constraint (\ref{primal_sub1}). Then the Lagrange function can  be written as
$$\mathcal{L} (\bm{x}^{(i)}, \bm{y}, \bm{\mu}^{(i)}) = f(\bm{x}^{(i)}, \bm{y}) + {\bm{\mu}^{(i)}}^T\bm{G}(\bm{x}^{(i)}, \bm{y}).$$
On the other hand, if Problem (\ref{Mprimal}) is infeasible, we add the iteration index $i$ into the infeasible primal problem index set, $\mathcal{I}$, as a new element and turn to an $l_1$-minimization feasibility-check problem as follows \cite{GBD}
\begin{eqnarray*}\label{Mfeasible}
	\hspace{10.2em}\min_{\{\bm{x}, \alpha\}} \alpha,\hspace{10.2em}
	\eqref{Mfeasible_obj} \label{Mfeasible_obj}
\end{eqnarray*}
\begin{subequations}
	subject to
	\vspace{-1em}
	\begin{align}
	\bm{G}(\bm{x}, \bm{y}^{(i-1)}) \leq \alpha,\label{feasible_sub1}
	\end{align}
	\vspace{-2em}
	\begin{align}
	\bm{x} \in \mathcal{X},\label{feasible_sub2}
	\end{align}
	\vspace{-2em}
	\begin{align}
	\alpha \geq 0.\label{feasible_sub3}
	\end{align}
\end{subequations}
Problem (\ref{Mfeasible}) is named as feasibility-check problem because its result reflects the feasibility of Problem (\ref{Mprimal}). Specifically, $\alpha=0$ means a feasible point of Problem (\ref{Mprimal}) has been found; otherwise, Problem (\ref{Mprimal}) is infeasible. As mentioned above, we only turn to Problem (\ref{Mfeasible}) when Problem (\ref{Mprimal}) is infeasible. Therefore, $\alpha=0$ will not happen in the GBD algorithm. By solving the feasibility-check problem, we can get the Lagrange multiplier vector $\bm{\lambda}^{(i)}$  for constraints (\ref{feasible_sub1}). The Lagrange function resulting from the infeasible primal problem is defined as 
$$\bar{\mathcal{L}} (\bm{x}^{(i)}, \bm{y}, \bm{\lambda}^{(i)}) =  {\bm{\lambda}^{(i)}}^T(\bm{G}(\bm{x}^{(i)}, \bm{y})-\alpha).$$ 
Note that an upper bound is only obtained from a feasible primal problem.

As for the master problem,  it is obtained based on nonlinear convex duality theory. Considering the Variant 2 of GBD (V2-GBD) \cite{GBD} without loss of generality, a relaxed master problem is given by
\vspace{-1em}
\begin{eqnarray*}\label{Mmaster}
	\hspace{10.5em}\min_{\{\bm{y},\eta\}} \eta, \hspace{10em}
	\eqref{Mmaster_obj} \label{Mmaster_obj}
\end{eqnarray*}
\begin{subequations}
	subject to
	\vspace{-1em}
	\begin{align}
	\bm{y} \in \mathcal{Y},\label{master_sub1}
	\end{align}
	\vspace{-2em}
	\begin{align}
	\eta \geq  \mathcal{L} (\bm{x}^{(j)}, \bm{y}, \bm{\mu}^{(j)}),  \forall j \in \mathcal{F}, \bm{\mu}^{(j)} \succeq \bm{0},  \label{master_sub2}
	\end{align}
	\vspace{-2em}
	\begin{align}
	0 \geq \bar{\mathcal{L}} (\bm{x}^{(j)}, \bm{y}, \bm{\lambda}^{(j)}),  \forall  j \in \mathcal{I},    \label{master_sub3}
	\end{align}
\end{subequations}
where constraints (\ref{master_sub2})  and (\ref{master_sub3}) are referred to as the optimality and feasibility cuts, respectively. Problem  (\ref{Mmaster}) is an MILP problem and can be solved by various MILP solvers, such as CPLEX\footnote{https://www.ibm.com/analytics/cplex-optimizer}. Note that CPLEX can be also used for MINLP problems. However, it can only solve the mixed integer second-order cone programming and mixed integer quadratic or quadratically constrained programming  problems, which only cover a small part of the resource allocation problems in wireless networks. The relaxed master problem provides a lower bound, ${{\eta}^*}^{(i)}$, to the original Problem (\ref{Mresource}) and its solution, $\bm{y}^{(i)}$, is used to generate the primal problem in the next iteration. The whole procedure of GBD is summarized in Table \ref{A1}. 

\begin{table}[h]
	\vspace{-1em}
	\setlength{\abovecaptionskip}{-2pt}
	\setlength{\belowcaptionskip}{-6pt}
	\caption{Generalized Benders Decomposition }
	\vspace{-1em}
	\label{A1}
	\begin{algorithm}[H]
		\caption{Generalized Benders Decomposition }
		{\small
			\begin{algorithmic}[1]
				\State \textbf{initialization}
				\State \quad Set iteration index, $i=0$.
				\State \quad Set maximum iteration number, $M$.
				\State \quad Set tolerance, $\Delta$.
				\State \quad Select an initial value for $\bm{y}^{(0)}$.
				\State \quad $UBD^{(0)} = \infty$, $LBD^{(0)}= -\infty$.
				\While {$|(UBD^{(i)}-LBD^{(i)})/LBD^{(i)}| > \Delta$ and $i<M$}
				\State $i=i+1$.
				\State Solve the primal problem (\ref{Mprimal}) by fixing $\bm{y}$ as $\bm{y}^{(i-1)}$.
				\If {the primal problem is feasible}
				\State Obtain the optimal solution $\bm{x}^{(i)}$.
				\State Obtain $\mathcal{L} (\bm{x}^{(i)}, \bm{y}, \bm{\mu}^{(i)})$ and an optimality cut $C^{(i)}$.
				\State Set $UBD^{(i)} = \min (UBD^{(i-1)},f(\bm{x}^{(i)}, \bm{y}^{(i-1)}) )$.
				\Else
				\State Solve the feasibility-check problem (\ref{Mfeasible}).
				\State Obtain $\bar{\mathcal{L}} (\bm{x}^{(i)}, \bm{y}, \bm{\lambda}^{(i)})$ and a feasibility cut $C^{(i)}$.
				\EndIf
				\State Add  $C^{(i)}$ into the relaxed master problem (\ref{Mmaster}) .
				\State Solve the relaxed master problem (\ref{Mmaster}) to obtain ${\eta}^*$ and $\bm{y}^{(i)}$.
				\State Set $LBD^{(i)} ={\eta}^*$.
				\EndWhile
		\end{algorithmic}}
	\end{algorithm}
\vspace{-3em}
\end{table}

\subsection{Multi-cut GBD}
A direct implementation of the classical GBD may require excessive computing time and memory. Many works have been dedicated to exploring ways to improve the convergence speed of the algorithm by reducing the required time for each iteration or the number of required iterations.  The former goal can be achieved by improving the procedure used to solve the primal and relaxed master problems at each iteration, which needs to be designed according to specific optimization problems. On the other hand, the latter goal can be achieved by improving the quality of the generated cuts, which is applicable for all optimization problems. In the following, we introduce an acceleration method of this kind named multi-cut GBD \cite{multicut}, which is the basis of our ML-based acceleration method to be introduced in Section III.

The key idea of multi-cut GBD is generating a number of cuts at each iteration to reduce the number of required iterations. The detailed procedure of the multi-cut GBD is summarized in Table \ref{A2}. The main difference between Algorithms \ref{A1} and \ref{A2} is that a set $\mathscr{S}$ of discrete solutions, instead of only one solution, is obtained while solving the relaxed master problem (\ref{Mmaster}). The set $\mathscr{S}$ includes both optimal and suboptimal solutions for Problem (\ref{Mmaster}). Generally, the size of $\mathscr{S}$ is set as a given constant, $S$. To get $\mathscr{S}$, we first use MILP solvers to get all feasible solutions for Problem (\ref{Mmaster}). Then we add the solution with the smallest gap between its objective value, $\eta$, and the optimum value, $\eta^*$, one by one into  $\mathscr{S}$ until $|\mathscr{S}|=S$. Note that $S$ should not be set large. Otherwise, there may not exist enough feasible solutions. If the number of existing solutions is less than $S$, we will collect all existing solutions as the set $\mathscr{S}$. In this way, we can get $|\mathscr{S}|$ primal problems by fixing the discrete variables, $\bm{y}$, in the Problem (\ref{Mprimal}) as the solutions in set $\mathscr{S}$. By solving the $|\mathscr{S}|$ primal problems, $|\mathscr{S}|$ cuts can be obtained and added to the relaxed master problem (\ref{Mmaster}). Then Problem (\ref{Mmaster}) is solved and a new set $\mathscr{S}$ is obtained, which is used to generate new $|\mathscr{S}|$ primal problems in the next iteration. 

According to \cite{multicut}, the multi-cut GBD does not change the optimal solution of the GBD algorithm for convex MINLP problems. The multiple cuts generated in the multi-cut GBD can improve the obtained lower bounds when solving the relaxed master problem. Therefore, the total number of required iterations  decreases and fewer relaxed master problems need to be solved as well. On the other hand, more primal problems need to be solved in multi-cut GBD than in classical GBD. Fortunately, the $|\mathscr{S}|$ primal problems are independent and can be solved in parallel while implementing multi-cut GBD, which will not bring additional overhead in terms of computational time if parallel computing is available. Overall, the total computational time can be expected to decrease when the multi-cut GBD is deployed, according to \cite{multicut}.
\begin{table}
	\vspace{-2em}
	\setlength{\abovecaptionskip}{-2pt}
	\setlength{\belowcaptionskip}{-6pt}
	\caption{Multi-cut Generalized Benders Decomposition }
	\vspace{-1em}
	\label{A2}
	\begin{algorithm}[H]
		\caption{Multi-cut Generalized Benders Decomposition }
		{\small
			\begin{algorithmic}[1]
				\State \textbf{initialization}
				\State \quad Set iteration index, $i=0$.
				\State \quad Set maximum iteration number, $M$.
				\State \quad Set tolerance, $\Delta$.
				\State \quad Select the initial solution set as $\mathscr{S}=\{\bm{y}^{(0)}\}$.
				\State \quad $UBD^{(0)} = \infty$, $LBD^{(0)}= -\infty$.
				\While {$|(UBD^{(i)}-LBD^{(i)})/LBD^{(i)}| > \Delta$ and $i<M$}
				\State $i=i+1$, $s=0$.
				\While {$|\mathscr{S}| \neq 0$}
				\State $s=s+1$.
				\State $\bm{y}_s \leftarrow$ pop out the best solution from $\mathscr{S}$.
				\State Solve the primal problem (\ref{Mprimal}) by fixing $\bm{y}$ as $\bm{y}_s$.
				\If {the primal problem is feasible}
				\State Obtain the optimal solution $\bm{x}^{(i)}$.
				\State Obtain $\mathcal{L} (\bm{x}^{(i)}, \bm{y}, \bm{\mu}^{(i)})$ and an optimality cut $C_s^{(i)}$.
				\State Set $UBD^{(i)} = \min (UBD^{(i-1)},f(\bm{x}^{(i)}, \bm{y}_s) )$.
				\Else
				\State Solve the feasibility-check problem (\ref{Mfeasible}).
				\State Obtain $\bar{\mathcal{L}} (\bm{x}^{(i)}, \bm{y}, \bm{\lambda}^{(i)})$ and a feasibility cut $C_s^{(i)}$.
				\EndIf
				\State Add $C_s^{(i)}$ into the relaxed master problem (\ref{Mmaster}) .
				\EndWhile
				\State Solve problem (\ref{Mmaster}) to obtain the optimal ${\eta}^*$ and a set  $\mathscr{S}$ of \indent \hspace{-1em} multiple solutions for  $\bm{y}$.
				\State Set $LBD^{(i)} ={\eta}^*$.
				\EndWhile
		\end{algorithmic}}
	\end{algorithm}
\vspace{-2em}
\end{table}

However, more cuts are added to the relaxed master problem at each iteration in the multi-cut GBD compared with the classical GBD. Given the fact that the relaxed master problem will be more time-consuming as more cuts are added, the main overhead of multi-cut GBD is the average time consumed by solving the relaxed master problem. Note that, among all the cuts generated at each iteration in multi-cut GBD, some cuts do not contribute much change on the lower bound but would increase the complexity of the relaxed master problem. We name these cuts as useless cuts and the others as useful cuts. Actually, only useful cuts need to be added into the relaxed master problem at each iteration. If we can distinguish between useful and useless cuts and only add useful cuts to the relaxed master problem at each iteration, the average complexity of the relaxed master problems in multi-cut GBD will decrease, which further accelerates the whole algorithm.  According to \cite{bdsurvey}, there is no need  to optimally solve Problem (\ref{Mmaster}) at each iteration to produce cuts for the convex MINLP problems. Optimality can be guaranteed even though we use the cuts generated from sub-optimal solutions of Problem (\ref{Mmaster}). Therefore, our proposed methods can still achieve optimality for the convex MINLP problems by only adding useful cuts to the relaxed master problem. In the following, we propose to use ML techniques to achieve this goal.

\section{Machine Learning Based GBD}
As mentioned above, we aim to accelerate GBD by distinguishing between useful and useless cuts. In this section, we first provide an overview of the ML approach. Specifically, we utilize two ML techniques to deal with the acceleration task. Then we design features, introduce data collection and label definition, and propose the training and testing processes for theses two ML techniques, respectively. Finally, we analyze the computational complexity and space complexity of the proposed methods.

\subsection{Overview of the Machine Learning Approach}
Clearly, the definition of useful and useless cuts is vague up to this point. Actually, there is no reliable way to identify useless cuts \cite{nodefine}. If we use the vague terms to label a cut, we can only quantify the degree to which the cut satisfies the concept of  useful or useless with a membership value in $[0,1]$. The membership value corresponds to some continuous performance indicators that can be directly collected by running the GBD algorithm.  Given that our ultimate goal is to only add useful cuts to the relaxed master problem, the continuous membership values need to be transformed into the binary labels. The transformation rule is handcrafted and heuristic, which will induce noise in the binary label of each cut. Generally, noisy labels have a significant impact on performance and need to be carefully dealt with. Therefore, we propose to utilize two different ML techniques, i.e., classification and regression, to deal with this acceleration task. Classification and regression are two classical techniques in ML. The task of classification is predicting a discrete label and that of regression is predicting a continuous quantity. Moreover, the algorithms and evaluation metrics of them are also different.  In the following, we introduce how to solve the acceleration task using these two ML techniques, respectively. We also analyze and compare their advantages and disadvantages in detail using the simulation results in Section IV.

\subsection{``Cut Classifier" Machine Learning Algorithm}
As mentioned above, classification aims to predict the discrete labels. Therefore, by using the classification technique, we directly train a model, i.e., a cut classifier, to predict the binary label of each cut. 
\subsubsection{Feature Design}
Feature design is very important for cut classifier training. To achieve better performance in the ML based GBD algorithm, we need to dig out proper features that are closely related to the generated cuts. Given that our proposed method is expected to be used in a large class of mixed integer wireless resource allocation problems, the proposed features are supposed to be invariant in different problems, i.e., to be problem-independent. Therefore, we only focus on the structure of GBD and do not take  into consideration the specific features of resource allocation problems and wireless networks. In this way, the designed features consist of the following five categories.
\begin{itemize}[leftmargin=3mm]
	\item Cut optimality indicator: 
	As mentioned in Section II, cuts are divided into two kinds: optimality cut and feasibility cut.  We set the cut optimality indicator as 1 for the optimality cut and 0 for the feasibility one.  The optimality cut is derived from the feasible primal problem (\ref{Mprimal}), while the feasibility cut is derived from the feasibility-check problem  (\ref{Mfeasible}).  As mentioned before, the upper bound is obtained only from the feasible primal problem while the  feasibility-check problem does not contribute to the convergence. Therefore, the optimality cut is preferred than the feasibility cut.
	
	\item Cut violation:
	Cut violation has been first proposed in \cite{svmgbd} for Benders decomposition. It can also be used in this work with some modifications. Cut violation means how large the feasible region of the relaxed master problem is cut off by adding a cut. For a cut generated at the current solution ($\bm{x}^{(i)}, \bm{y}^{(i-1)},\eta^{(i-1)}$), the cut violation is defined as $\mathcal{L} (\bm{x}^{(i)}, \bm{y}^{(i-1)}, \bm{\mu}^{(i)}) - \eta^{(i-1)}$  if the cut is an optimality cut. And for the feasibility cut, its cut violation is defined as $\bar{\mathcal{L}} (\bm{x}^{(i)}, \bm{y}^{(i-1)}, \bm{\lambda}^{(i)})$. Larger cut violation means that the new feasible region after adding the cut is far from the current solution, which is expected to bring about larger change on the lower bound and thus is preferred.

	\item Cut repeat:
	A cut may be generated more than once during the multi-cut GBD algorithm. As mentioned above, we collect both optimal and sub-optimal solutions for the relaxed master problem at each iteration during the multi-cut GBD algorithm. A specific $\bm{y}$ may be the sub-optimal solution for the relaxed master problems in different iterations, thus its corresponding cut might be generated more than once. For a cut generated at iteration $i$, its cut repeat is defined as how many times this cut has been generated until the $i$-th iteration. Cut repeat is a very important feature. Repeatedly adding existing cuts may lead to redundancy without any contribution to the change of lower bound.

	\item Cut depth: 
	For a cut generated at iteration $i$, its cut depth is $i$. Because the objective value of the relaxed master problem, $\eta$, increases in a decreasing speed over iterations, we cannot expect the change amount of $\eta$ brought by the cuts generated at different iterations to be equally large. On the other hand, usually fewer cuts need to be added to the relaxed master problem over iterations. Therefore, cut depth is also a necessary feature for the generated cuts. 
	
	\item Cut order: While solving the relaxed master problem in multi-cut GBD, multiple solutions for the discrete variables, $\bm{y}$, are generated. Each solution corresponds to a cut.  As mentioned in Section II.C, at each iteration, we first generate all feasible solutions for the relaxed master problem, and then add the solution with the smallest gap value, $\eta-\eta^*$, into the solution set $\mathscr{S}$ one by one until $|\mathscr{S}|=S$ or all feasible solutions are already added into $\mathscr{S}$. In this way, different solution is added into $\mathscr{S}$ in a different order, which is also defined as its corresponding cut's order. To some extent, cut order can reflect the quality of the cut.
\end{itemize}

With the above five features, each cut is described by a five-dimensional vector, which is then input to the cut classifier.

\subsubsection{Data Collection  and Label Definition}
When we focus on a specific resource allocation problem, we can generate a training problem set, $\mathscr{Q}$, which includes different problem instances. Then, we propose to run a modified GBD algorithm on each problem instance in $ \mathscr{Q}$ to collect the generated cuts as the training samples. Specifically, a set $\mathscr{S}$ of multiple discrete solutions are obtained from the relaxed master problem at each iteration of the modified GBD. Then we only randomly choose one solution from $\mathscr{S}$  and fix the discrete variables, $\bm{y}$, as the selected solution to get the primal problem (\ref{Mprimal}). We solve this primal problem and get a cut. Finally, we add the cut to the relaxed master problem and solve it. The above process is different from the multi-cut GBD, where multiple cuts are derived at each iteration. It is also different from the classical GBD, where the cut at each iteration is always derived from the optimal solution of the relaxed master problem and lacks variety.
\begin{table}
	\vspace{-2em}
	\setlength{\abovecaptionskip}{-2pt}
	\setlength{\belowcaptionskip}{-6pt}
	\caption{Data Collection Method for Cut Classifier}
	\vspace{-1em}
	\label{A3}
	\begin{algorithm}[H]
		\caption{Data Collection Method for Cut Classifier}
		{\small
			\begin{algorithmic}[1]
				\State \textbf{initialization}
				\State \quad Set training dataset: $\mathscr{T}=\emptyset$.
				\For {problem instance $Q$ \textbf{in}  $\mathscr{Q}$}
				\State \textbf{initialization}
				\State \quad Set iteration index, $i=0$.
				\State \quad Set maximum iteration number, $M$.
				\State \quad Set tolerance, $\Delta$.
				\State \quad Select the initial solution set as $\mathscr{S}=\{\bm{y}^{(0)}\}$.
				\State \quad $UBD^{(0)} = \infty$, $LBD^{(0)}= -\infty$.
				\While {$|(UBD^{(i)}-LBD^{(i)})/LBD^{(i)}| > \Delta$ and $i<M$}
				\State $i=i+1$.
				\State $\bm{y}^{(i-1)}\leftarrow$ randomly pop out a solution from $\mathscr{S}$.
				\State Solve the primal problem (\ref{Mprimal}) by fixing $\bm{y}$ as $\bm{y}^{(i-1)}$.
				\If {the primal problem is feasible}
				\State Obtain the optimal solution $\bm{x}^{(i)}$.
				\State Obtain $\mathcal{L} (\bm{x}^{(i)}, \bm{y}, \bm{\mu}^{(i)})$ and  an optimality cut $C^{(i)}$.
				\State Set $UBD^{(i)} = \min (UBD^{(i-1)},f(\bm{x}^{(i)}, \bm{y}^{(i-1)}) )$.
				\Else
				\State Solve the feasibility-check problem (\ref{Mfeasible}).
				\State Obtain $\bar{\mathcal{L}} (\bm{x}^{(i)}, \bm{y}, \bm{\lambda}^{(i)})$ and a feasibility cut $C^{(i)}$.
				\EndIf
				\State Add $C^{(i)}$ into the relaxed master problem (\ref{Mmaster}) .
				\State Solve problem (\ref{Mmaster}) to obtain the optimal ${\eta}^*$ and a set  $\mathscr{S}$ \indent \indent \hspace{-1.5em} of multiple solutions for  $\bm{y}$.
				\State Set $LBD^{(i)} ={\eta}^*$.
				\State $CI^{(i)} = LBD^{(i)} - LBD^{(i-1)}$.
				\If {$i>1$}
				\If {$CI^{(i)}/CI^{(i+1)}>\theta$}
				\State $\mathscr{T} \leftarrow \{Feature(C^{(i-1)}), 1\}$.
				\Else
				\State $\mathscr{T} \leftarrow \{Feature(C^{(i-1)}), 0\}$.
				\EndIf
				\EndIf
				\EndWhile
				\State$\mathscr{T} \leftarrow \{Feature(C^{(i-1)}), 1\}$.
				\EndFor
				\State  \textbf{return} training set $\mathscr{T}$.
		\end{algorithmic}}
	\end{algorithm}
\vspace{-1em}
\end{table}

Because we want to learn the cut classifier in a supervised manner, both the features and the label for each cut should be collected during the data collection process. As mentioned above, the binary label of each cut, denoted as $BL$, cannot be directly collected by running the GBD algorithm. It is  transformed from some continuous performance indicators with handcrafted rules. In our work, we assign $BL$ as $1$ for a useful cut and $0$ for a useless one. Here, we denote the change amount of the objective value of the relaxed master problem, $\eta$, before and after adding a certain cut at the $i$-th iteration as $CI^{(i)}$. We use $CI^{(i)}$ as the continuous performance indicator of the cut derived at the $i$-th iteration. As mentioned before, the objective value of the relaxed master problem increases in a decreasing speed over the iterations. Therefore, we prefer the cut that can bring a large $CI$ as compared to the cut in the next iteration. In this way, the binary label of the cut derived at the $i$-th iteration, $BL^{(i)}$, is assigned as 1 if $CI^{(i)}/CI^{(i+1)}>\theta$, and as 0 otherwise.  Meanwhile, the binary label of the cut derived at the last iteration is always set as 1. In summary, the details of the data collection method for cut classifier are shown in Table \ref{A3}.  

\begin{figure}
	\vspace{-1em}
	\centering
	\includegraphics[width=0.6\linewidth, height=0.11\textheight]{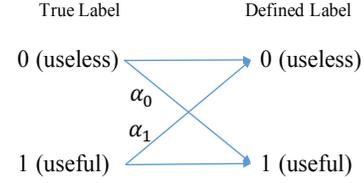}
	\vspace{-1em}
	\caption{Relations between the true labels and the defined labels.}
	\label{fig:error_relation}
	\vspace{-2em}
\end{figure}

As mentioned in Section III.A, the handcrafted transformation rule is heuristic and will induce noise in the binary labels. The relation between the true labels and the defined labels can be illustrated as Fig. \ref{fig:error_relation}.  To alleviate the impact of noise, choosing an appropriate threshold, $\theta$, is very important. We initially set threshold $\theta$ as $1$ and adjust it according to the output performance. According to  \cite{noiselabel2}, the existence of noise will not change the misclassification probability if $\alpha_0 = \alpha_1$. It suggests that we can avoid the influence of noise by achieving a balance between $\alpha_0$ and $\alpha_1$. Clearly, a larger threshold means fewer cuts are defined as  useful, which leads to a larger $\alpha_1$ and a smaller $\alpha_0$. Meanwhile, a smaller threshold leads to the opposite result. Based on these facts, we can carefully adjust  $\theta$ to achieve the balance between $\alpha_0$ and $\alpha_1$ and alleviate the impact of noise although we do not know the exact values of error rates. If the accuracy of the learned classifier is not satisfactory because of $\alpha_0 \neq \alpha_1 $, we may try a larger $\theta$ and a smaller one at the same time since we do not know which error is bigger. One of the new thresholds will decrease  $|\alpha_0-\alpha_1|$, which brings about better performance. Otherwise, the one adopted now is already the best threshold and we need to try other methods for performance improvement. 

\subsubsection{Training Process}
After collecting the training dataset $\mathscr{T}$ using Algorithm \ref{A3}, we can adopt some well-known classification techniques, such as support vector machine (SVM) \cite{svm}, linear discriminant analysis (LDA) \cite{lda}, and logistic regression \cite{lr}, to learn the cut classifier. Given the fact that the number of the useless cuts is generally greater than that of the useful cuts, the collected training dataset $\mathscr{T}$ is imbalanced, which will influence the performance of the learned cut classifier. Therefore, we use the undersampling technique \cite{undersample} on the collected dataset $\mathscr{T}$ before training. This technique uses a subset of the majority class for training. Since many majority class samples are ignored, the training set becomes more balanced and is denoted as $\mathscr{\hat{T}}$. Then we use the aforementioned ML classification techniques with sampled training dataset $\mathscr{\hat{T}}$ and plot the receiver operating characteristic (ROC) curves to evaluate the diagnostic ability of learned cut classifiers with different  classification techniques. We choose the best one for the following testing process.

The overhead of the data collection process and the training process consists of the following three parts: the time consumed by collecting the training dataset $\mathscr{T}$ from the  training problem set $\mathscr{Q}$, the memory consumed by storing  $\mathscr{T}$, and the time consumed by learning the cut classifier from $\mathscr{T}$. As mentioned above, the classification techniques we adopt are all from the classical ML field. They only need a small number of training samples to get good models and converge  fast, which is different from the deep learning technique. In this way, the overhead of collecting and storing $\mathscr{T}$ is modest, and the time consumed by learning the cut classifier is also negligible.

We shall note that the training process of our proposed method is offline. Moreover, the proposed method has generalization ability on different aspects and a learned cut classifier can be used for more than one applications, which will be validated by the simulation results in Section IV. In this way, the overhead of the training process may be amortized in practice over different applications. In summary, the overhead of the data collection process and the  training process will not bring negative effects on the performance of our proposed method and we will not take this overhead into consideration for the following testing process.

\subsubsection{Testing Process}
The testing process is very similar to the multi-cut GBD in Table \ref{A2}. The only difference is that we need to use the learned cut classifier to determine whether the cut is useful before adding the generated cuts into the relaxed master problem. We add it to the relaxed master problem only if the cut is classified as useful. However, if all the $|\mathscr{S}|$ generated cuts are labeled as useless cuts by the cut classifier, we switch to the classical GBD and add the cut generated by the optimal solution $\bm{y}^*$. The detailed testing process is summarized in Table \ref{A4}.

\begin{table}[h]
	\vspace{-1em}
	\setlength{\abovecaptionskip}{-2pt}
	\setlength{\belowcaptionskip}{-6pt}
	\caption{Testing Process with Cut Classifier $\pi_c$ or Cut Regressor $\pi_r$}
	\vspace{-1em}
	\label{A4}
	\begin{algorithm}[H]
		\caption{Testing Process with Cut Classifier $\pi_c$ or Cut Regressor $\pi_r$}
		{\small
			\begin{algorithmic}[1]
				\State \textbf{initialization}
				\State \quad Set iteration index, $i=0$.
				\State \quad Set maximum iteration number, $M$.
				\State \quad Set tolerance, $\Delta$.
				\State \quad Select the initial solution set as $\mathscr{S}=\{\bm{y}^{(0)}\}$.
				\State \quad $UBD^{(0)} = \infty$, $LBD^{(0)}= -\infty$.
				\While {$|(UBD^{(i)}-LBD^{(i)})/LBD^{(i)}| > \Delta$ and $i<M$}
				\State $i=i+1$, $s=0$, $n_{cut} = 0$.
				\While {$|\mathscr{S}| \neq 0$}
				\State $s=s+1$.
				\State $\bm{y}_s \leftarrow$ pop out the best solution from $\mathscr{S}$.
				\State Solve the primal problem (\ref{Mprimal}) by fixing $\bm{y}$ as $\bm{y}_s$.
				\If {the primal problem is feasible}
				\State Obtain the optimal solution $\bm{x}^{(i)}$.
				\State Obtain $\mathcal{L} (\bm{x}^{(i)}, \bm{y}, \bm{\mu}^{(i)})$ and get an optimality cut $C_s^{(i)}$.
				\State Set $UBD^{(i)} = \min (UBD^{(i-1)},f(\bm{x}^{(i)}, \bm{y}_s) )$.
				\Else
				\State Solve the feasibility-check problem (\ref{Mfeasible}).
				\State Obtain $\bar{\mathcal{L}} (\bm{x}^{(i)}, \bm{y}, \bm{\lambda}^{(i)})$  and get a feasibility cut $C_s^{(i)}$.
				\EndIf
				\EndWhile
				\While {$s>0$}
				\State Using $\pi_c$ or $\pi_r$ to decide whether $C_s^{(i)}$ is useful. 
				\State If useful, add it into the relaxed master problem (\ref{Mmaster}).
				\State $n_{cut} = n_{cut} + 1$, $s = s - 1$.
				\EndWhile
				\If {$n_{cut}=0$}
				\State Add $C_1^{(i)}$ into the relaxed master problem (\ref{Mmaster}).
				\EndIf
				\State Solve problem (\ref{Mmaster}) to obtain the optimal ${\eta}^*$ and a set  $\mathscr{S}$ of \indent \hspace{-1em} multiple solutions for  $\bm{y}$.
				\State Set $LBD^{(i)} ={\eta}^*$.
				\EndWhile
		\end{algorithmic}}
	\end{algorithm}
	\vspace{-3em}
\end{table}

\subsection{``Cut Regressor" Machine Learning Algorithm}
Different from the cut classifier that directly predicts the binary label,  the task of regression is predicting a continuous quantity. Therefore, by using the regression technique, we train a model, i.e., a cut regressor, to predict the continuous performance indicator of each cut.
\subsubsection{Feature Design}
In Section III.B, five features closely related to the generated cuts are proposed, which are invariant for different problems. We still use these five features for training cut regressor.

\subsubsection{Data Collection and Label Definition}
For the ``cut regressor" ML problem, we propose to collect training dataset by directly running each problem instance in the training problem set, $\mathscr{Q}$, with the multi-cut GBD in Table \ref{A2}. Since the goal of cut regressor is learning the continuous performance indicators instead of the binary labels of each cut, the label for each collected data should be the continuous performance indicator. Different from the data collection process for cut classifier, multiple training samples are generated at each iteration. In this case, we cannot use $CI^{(i)}/CI^{(i+1)}$ as the continuous performance indicators any more, because samples generated at the same iteration are related and we need to take their relations into consideration. To deal with this issue, we order the cuts generated at the same iteration using the same method mentioned in Section III.B and propose two new indicators. The first one is the change amount of the objective value of the relaxed master problem, $\eta$,  before and after adding the first $s$ cuts at the $i$-th iteration, denoted as $ACI_s^{(i)}$. $ACI_s^{(i)}$ reflects the cumulative influence of the first $s$ cuts and does not decrease with $s$. The second one is the change amount of the objective value of the relaxed master problem, $\eta$,  before and after adding all $|\mathscr{S}|$ cuts at the $i$-th iteration, denoted as $CT^{(i)}$. To get normalized labels that are preferred in ML, we introduce a new variable $CR_s^{(i)}$ as the performance indicator for the $s$-th cut at the $i$-th iteration, which is defined as the ratio between $ACI_s^{(i)}$ and $CT^{(i)}$, i.e., $CR_s^{(i)}=ACI_s^{(i)}/CT^{(i)}$.  As mentioned above, $ACI_s^{(i)}$ does not decrease with $s$. Therefore, $CR_s^{(i)}$ also does not decrease with s and its' value varies from 0 to 1. During the data collection process, we collect aforementioned five features and the proposed performance indicator, $CR_s^{(i)}$, as the label for the $s$-th cut at the $i$-th iteration. 

\subsubsection{Training Process}
After collecting the training dataset, we can adopt some well-known regression techniques, such as extra tree \cite{extratree}, linear regression \cite{linearegression}, ridge regression \cite{ridge}, lasso regression \cite{lasso} and SVM regression \cite{svm}, to learn the cut regressor. Note that there is no data imbalance issue for regression problems and the undersampling process is not needed before training. To evaluate how well the model fits the data, we check the R-squared score \cite{r2} of each regression model.  Suppose that we have a training set $\mathscr{T}= \{(\bm{u}^{(n)},z^{(n)}); n = 1, ..., N\}$ consisting of $N$ training samples, where $\bm{u}^{(n)}$ is the vector of input features and $z^{(n)}$ is the label. Then the R-squared score is defined as 
$$R^2 = 1- \frac{\sum_{n=1}^{N}(\hat{z}^{(n)}-z^{(n)})^2}{\sum_{n=1}^{N}(\bar{z}-z^{(n)})^2},$$
where $\hat{z}^{(n)}$ is the predicted result of sample $(\bm{u}^{(n)},z^{(n)})$ and $\bar{z} = \sum_{n=1}^N z^{(n)}/N$. The R-squared score indicates the percentage of the response variable variation that is explained by the regression model, and therefore it is always between 0 and 1. Generally, the larger the R-squared score is, the better the regression model fits the data.  Therefore, we choose the regression model with the largest  R-squared score for the following testing process. In addition, the overhead of the data collection process and the training process for the cut regressor is similar to that of the cut classifier, which is negligible and will not be considered during the following testing process.

\subsubsection{Testing Process and Handcrafted Label Transformation Rule}
\begin{figure}[h]
	\vspace{-1em}
	\centering
	\includegraphics[width=0.7\linewidth, height=0.2\textheight]{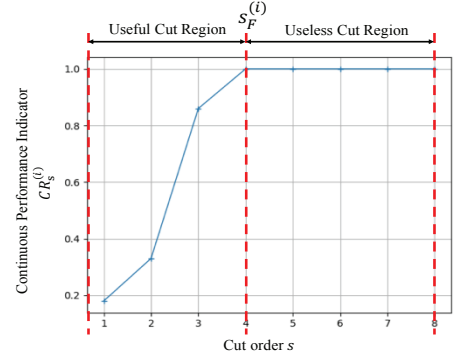}
	\vspace{-1em}
	\caption{An example of the relation between $CR_s^{(i)}$ and $s$.}
	\label{fig:CR}
	\vspace{-0.5em}
\end{figure}

The testing process of the cut regressor is also very similar to that of the cut classifier. The difference is that we can only get the predicted continuous indicator for each cut instead of directly knowing whether the cut is useful before adding the generated cuts into the relaxed master problem.  To transform the continuous indicator into the binary label $BL \in \{0,1\}$ during the testing process, we propose the following rule:
\begin{align}
BL_s^{(i)} =\left\{\begin{array}{rcl}
0, &        & s>1 \ \text{and} \ \hat{CR}_s^{(i)}=\hat{CR}_{s-1}^{(i)}=1,\\
1,&       &\text{otherwise},\\
\end{array} \right. \label{tansrule}
\end{align}
where $\hat{CR}_s^{(i)}$ is the predicted continuous performance indicator for the $s$-th cut at the $i$-th iteration. This rule can be explained as follows. During the multi-cut GBD, we always first use the best solution from $\mathscr{S}$. Therefore, when $s=1$, the cut is generated from the optimal solution $\bm{y}^*$. Given that at least one cut should be added to the relaxed master problem, the cut with $s=1$ must be defined as a useful cut. Meanwhile, $CR_s^{(i)} \in[0,1]$ and it does not decrease with $s$. An example of 8-cut GBD is shown in Fig. \ref{fig:CR}, where  $CR_s^{(i)}$ increases with $s$ and quickly reaches 1. Define  $s_{F}^{(i)}$ as the critical point such that $CR_s^{(i)}=1$ when $s \geq s_F^{(i)}$. Then, for the cut $C_s^{(i)}$ with $s \leq s_{F}^{(i)}$, it contributes to the change of the objective function of the relaxed master problem and should be defined as a useful cut. We call the interval $[1,s_{F}^{(i)}]$ as the useful cut region. Meanwhile, the cut $C_s^{(i)}$ with $s > s_{F}^{(i)}$ does not contribute to the change of the objective function of the relaxed master problem and should be defined as a useless cut.  We call the interval $(s_{F}^{(i)},S]$ as the useless cut region. It is easy to verify that $\hat{CR}_s^{(i)}=\hat{CR}_s^{(i-1)}=1$ is an equivalent condition for the useless cut region.

With the predicted  continuous indicator and the label transformation rule in (\ref{tansrule}), we can determine whether a cut is useful. And only useful cuts are added to the relaxed master problem. Note that the situation where all the $|\mathscr{S}|$ generated cuts are labeled as useless cuts will not happen in the testing process for the cut regressor, because we always define the cut with $s=1$ as a useful one in the transformation rule. The detailed testing process is also summarized in Table \ref{A4}.

\subsection{Computational Complexity and Space Complexity Analysis}
In this subsection, we analyze the computational complexity and space complexity of our proposed method, respectively. 

The computational complexity analysis of GBD algorithm is still open in the literature. Given the NP-hardness of the considered MINLP problems, the worst-case computational complexity of the proposed algorithm still grows exponentially with the problem size, while the average computational complexity is usually unknown despite of some special cases. Therefore, one may pay more attention to testing the practical computational complexity of an accelerated GBD algorithm in general. The practical computational complexity is obtained by doing simulations, which is further discussed  in Section IV.

To analyze the space complexity, we notice that the most memory-consuming part of the GBD algorithm lies in the relaxed master problem because the cumulative number of cuts added to the relaxed master problem, denoted as $S_c$, increases over the iterations. Therefore, the space complexity of GBD is directly related to $S_c$. Furthermore, each cut is generated by sequentially solving the primal and the relaxed master problems, where the value of $\bm{y}$ from the last iteration and the value of $\bm{x}$ from the primal problem are needed to be stored sequentially. Therefore,  the space complexity of the GBD algorithm can be expressed as $O(S_c \cdot \max(n_1,n_2))$, where $n_1$ and $n_2$ are the dimensions of $\bm{x}$ and $\bm{y}$, respectively. Given $n_1$ and $n_2$, the multi-cut GBD  algorithm generally adds more cuts to the relaxed master problem than the classical one. Therefore, the multi-cut GBD algorithm has a larger $S_c$ and is more memory-consuming than the classical one. On the other hand, our proposed method can discard useless cuts. As suggested from the following simulation results in Section IV, the $S_c$ of our proposed method is close to that of the classical GBD, and much smaller than that of the multi-cut GBD. Therefore, our proposed method has a similar space complexity to the classical GBD and a much lower one compared with the unaccelerated multi-cut GBD algorithm.  

\section{Testing Results}
In this section, we test the performance of the proposed ML based GBD. We use a joint subcarrier and power allocation problem in D2D networks as an example. We first briefly introduce the system model and resource allocation problem. Then we present the testing results of the proposed method on this problem. We also pay  special attention to the generalization ability of the proposed method, which is difficult to meet but is preferred in practice.

\subsection{Resource Allocation Problem  in D2D Networks}
We adopt the system model and problem formulation in \cite{learntobranch}, where an uplink single-cell scenario with  $K$ cellular users (CUs) in a set $\mathscr{K}$ and $L$ D2D pairs in a set $\mathscr{L}$ is considered. Each CU connects to the evolved nodeB (eNB) with an orthogonal channel and each D2D pair transmits data by reusing the uplink channels of CUs. To optimize the overall data rate and achieve user fairness, a \emph{max-min} optimization problem is formulated. Specifically, the objective function is to maximize the minimum data rate among all  D2D pairs, where the transmit power vector of CUs,  $\bm{p^\text{C}}$, the transmit power vector of D2D pairs, $\bm{p^\text{D}}$, and the  discrete indicator vector of subcarrier allocation, $\bm{\rho}$, are to be determined. 

We consider the following constraints. First, the data rate of each CU is required to be no less than a given threshold, $R^\text{C}_{\min}$. Second, the power of individual D2D pair is constrained by an upper bound, $P^\text{D}_{\max}$.  Third, the power of each CU is also constrained by an upper bound, $P^\text{C}_{\max}$. Finally, each subcarrier can be reused by at most one D2D pair to limit the interference. In this way, the mathematical formulation of this MINLP problem is given by \cite{learntobranch}
\begin{eqnarray*}\label{Md2d}
	\hspace{6em}\max_{\{\bm{p^\text{C}}, \bm{p^\text{D}},\bm{\rho}\}} \min_{l\in \mathscr{L} } R_{l}^\text{D}(\bm{p^\text{C}},\bm{p^\text{D}},\bm{\rho}),\hspace{5.4em}
	\eqref{Md2d_obj} \label{Md2d_obj}
\end{eqnarray*}
\begin{subequations}
	subject to
	\vspace{-1em}
	\begin{align}
	\rho_{kl} \in \{0,1\}, \quad{\forall}k\in \mathscr{K}, l\in \mathscr{L},\label{d2d_sub1}
	\end{align}
	\vspace{-1em}
	\begin{align}
	\sum_{l\in \mathscr{L}}\rho_{kl}\leq 1, \quad{\forall}k\in \mathscr{K},\label{d2d_sub2}
	\end{align}
	\vspace{-1em}
	\begin{align}
	\sum_{k\in \mathscr{K}}\rho_{kl}p_{kl}^\text{D} \leq P^\text{D}_{\max}, \quad{\forall}l\in \mathscr{L},\label{d2d_sub3}	
	\end{align}
	\vspace{-1em}
	\begin{align}
	R_{k}^\text{C}(\bm{p^\text{C}},\bm{p^\text{D}},\bm{\rho})\geq R^\text{C}_{\min}, \quad{\forall}k\in \mathscr{K},\label{d2d_sub4}
	\end{align}
	\vspace{-2em}
	\begin{align}
	p_{k}^\text{C}\leq P^\text{C}_{\max}, \quad{\forall}k\in \mathscr{K},\label{d2d_sub5}
	\end{align}
\end{subequations}
where $R_{k}^\text{C}$  and $R_{l}^\text{D}$ are the data rates of CU $k$ and D2D pair $l$, respectively. They are given by
$$
R_{k}^\text{C}(\bm{p^\text{C}},\bm{p^\text{D}},\bm{\rho})= \log(1+\frac{p_k^\text{C}h_{k}^\text{CB}}{\sigma_N^2+\sum_{l\in \mathscr{L}} \rho_{kl}p_{kl}^\text{D}h_{l}^\text{DB}}),$$
$$
R_{l}^\text{D}(\bm{p^\text{C}},\bm{p^\text{D}},\bm{\rho})=   \sum_{k\in \mathscr{K}}\rho_{kl} \log(1+\frac{\rho_{kl}p_{kl}^\text{D}h_l^\text{D}}{\sigma_N^2+p_k^\text{C}h_{kl}^\text{CD}}), $$
where $\sigma_N^2$ is the power of the additive white Gaussian noise (AWGN), $\bm{h^\text{CD}}$, $\bm{h^\text{CB}}$, $\bm{h^\text{D}}$, and $\bm{h^\text{DB}}$ are the channel power gains between CU and the receiver of the D2D pair, CU and the eNB, the transmitter and the receiver of the D2D pair, and the transmitter of the D2D pair and the eNB, respectively.  Since the data rates of D2D users are independent of each other, it can be proven that Problem  (\ref{Md2d}) satisfies the convexity and linear separability mentioned in Section II.B. Detailed discussions of its convexity can be found in \cite{learntobranch} and omitted here to avoid redundancy. Note that the resource allocation problems in \cite{chen,MECO} are similar to  Problem (\ref{Md2d}), which suggests that Problem (\ref{Md2d}) is a suitable example to verify the effectiveness  of our proposed method.

\subsection{Simulation Setup}
\subsubsection{Parameters for D2D Network}
We consider a single-cell network with a radius of 500 m, where the eNB is located in the center and all the users are distributed uniformly in the cell. Some important network parameters are given in Table \ref{para} and more details can be found in  \cite{learntobranch}. 
\begin{table}
	\vspace{-2em}
	\scriptsize
	\caption{D2D Network Parameters}
	\vspace{-1em}
	\label{para}
	\centering
	\begin{tabular}{|c|c|}
		\hline
		Parameter & Value  \\
		\hline
		\hline
		Cell Radius & 500 m \\
		\hline
		Noise Spectral Density & -174 dBm/Hz\\
		\hline
		\tabincell{c}{Path Loss Model  for Cellular Links} & 128.1+37.6log(d[km])\\
		\hline
		\tabincell{c}{Path Loss Model for D2D Links} & 148+40log(d[km])\\
		\hline
		Shadowing Standard Deviation & 10 dB \\
		\hline
		\tabincell{c}{Maximum Transmitter Power of CU, $P^\text{C}_{\max}$} & 20 dBm \\
		\hline
		\tabincell{c}{Maximum Transmitter Power of D2D Pair, $P^\text{D}_{\max}$} & 20 dBm \\
		\hline
		Minimum Data Rate of CU, $R^\text{C}_{\min}$ & 2 bit/s/Hz\\
		\hline
	\end{tabular}
\vspace{-2em}
\end{table}

\subsubsection{Parameters for Three Kinds of GBD algorithms}
During the simulation, we compare the performance of classical GBD, multi-cut GBD, and ML based GBD. For all these three kinds of GBD algorithms, we set the convergence tolerance for the upper and lower bounds, $\Delta$, as 0.5\%, and the maximum iteration number, $M$, as 10,000. We use the CPLEX 12.9 to obtain the solutions of the relaxed master problems and get the multiple solutions by using solution pool and setting relevant parameters of CPLEX.  All other codes are implemented in python 3.6 except the solving process for the primal problem that is implemented in Matlab. The code is run on the Windows system with a 3.6 GHz CPU. In the following, we call the classical GBD as single-cut GBD, and the multi-cut GBD with $S=s$ as $s$-cut GBD.

To choose an appropriate value of $S$ for the multi-cut GBD, the data collection process, and the testing process, we test the performance of multi-cut GBD with different values of $S$. The results are summarized in Table \ref{S_influence}, where the average cut number  is defined as the ratio between the cumulative cut number and the needed iteration number.  As mentioned in Section II.C, there may not exist enough feasible solutions if $S$ is large. When the number of existing feasible solutions is smaller than $S$, we collect all existing solutions as the set $\mathscr{S}$ and $|\mathscr{S}|<S$.  Therefore, the average cut number is generally smaller than $S$. The multi-cut GBD algorithm with different $S$ may have similar average cut number and performance. In this case, a smaller $S$ is usually preferred, which is closer to the average cut number and better reflects the characteristics of the multi-cut GBD algorithm. On the other hand,  the cumulative time of multi-cut GBD first decreases and then fluctuates with the increase of $S$. Therefore, there will exist a critical point where the average cut number is close to $S$ and the cumulative time begins to fluctuate. The critical point corresponds to an appropriate value of $S$ for the baseline multi-cut GBD algorithm. Specifically, the critical point for our tested problem is $S=8$. Therefore, we choose $S=8$ for the following testing process and only compare the proposed method with 8-cut GBD. Note that the value of $S$  is usually pre-determined  and limited to a small range in practice \cite{multicut}. Moreover, finding the critical point of $S$ aims at choosing a baseline for fair comparison, which is not needed in practice and no overhead will  be incurred.

\begin{table*}
	\vspace{-2em}
	\scriptsize
	\caption{Performance of GBD Algorithms with Different Values of $S$}
	\vspace{-1em}
	\label{S_influence}
	\centering
	\begin{tabular}{|c|c|c|c|c|c|c|c|c|c|c|}
		\hline
		$S$&1 & 2&4&6&8&10&12&14&16&20 \\
		\hline
		\tabincell{c}{Needed Iteration Number}& 73.22& 43.85& 28.56& 25.00& 23.24& 21.54& 20.51&20.50&19.98&19.59\\
		\hline
		\tabincell{c}{Cumulative Cut Number} &73.22 & 86.68& 106.57&131.49& 152.37& 161.72&172.73&187.39&197.11&218.54 \\
		\hline
		\tabincell{c}{Average Cut Number} & 1.00& 1.98& 3.73& 5.26& 6.56& 7.51&8.42&9.14&9.86&11.16\\
		\hline
		Cumulative Time (s)& 2.97& 2.68&  2.12& 1.68& 1.50& 1.43& 1.41& 1.33& 1.36& 1.37 \\
		\hline
	\end{tabular}
\vspace{-1em}
\end{table*}

\subsection{Performance of Machine Learning Based GBD}
In this part, we focus on the performance of the proposed ML based GBD. We set the number of CUs, $K$, as 5, and that of D2D pairs, $L$, as 3. The training and testing problem sets both include 50 problem instances that are generated by randomly placing CUs and D2D pairs. Note that 50 problem instances correspond to thousands of cut samples and we report the average performance over the testing problem set.  

\begin{figure}
	\vspace{-1em}
	\centering
	\includegraphics[width=0.65\linewidth, height=0.2\textheight]{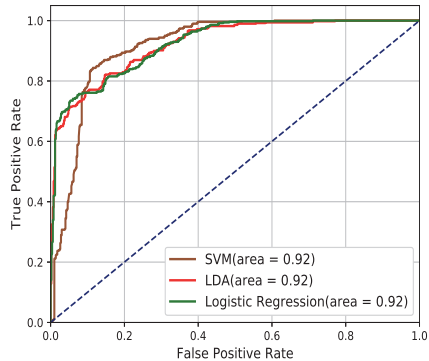}
	\vspace{-1em}
	\caption{ROC curves for different classification techniques.}
	\label{fig:roc}
	\vspace{-1em}
\end{figure}

As mentioned in Section III.B, we have adopted three well-known ML classification techniques to learn the classifier. During the training process, we set the class weights for useful and useless cuts as 2:1. The ROC curves of the learned cut classifiers with different techniques are shown in Fig. \ref{fig:roc}. We usually use the area under the ROC curve to measure how well a classifier is. The larger the area is, the better the classifier performs. From the figure, SVM, LDA, and logistic regression perform equally well, i.e., the areas under the ROC curves for all of them achieve 0.92. Thus, we simply use the cut classifier from SVM in the following test. 

As mentioned in Section III.C, we also have adopted five popular ML regression techniques to learn the regressor and used the R-squared scores to evaluate them. We summarize the R-squared scores for them in Table \ref{r2score}. Generally, regression model with a larger R-squared score is preferred.  Therefore, we use  the cut regressor from extra tree in the following test. From Table \ref{r2score}, linear regression and another two special linear regression models, i.e., ridge regression and lasso regression, have very small R-squared scores. The results suggest that the relations between the features and the continuous performance indicators are highly non-linear. 
\begin{table}
	\scriptsize
	\caption{R-squared Scores for Different Regression Techniques}
	\vspace{-1em}
	\label{r2score}
	\centering
	\begin{tabular}{|c|c|}
		\hline
		Regression Model& R-squared Score \\
		\hline
		Extra Tree & 0.99 \\
		\hline
		Linear Regression & 0.08\\
		\hline
		Ridge Regression& 0.08\\
		\hline
		Lasso Regression& 0.01\\
		\hline
		 SVM Regression &0.81\\
		 \hline
	\end{tabular}
	\vspace{-1em}
\end{table}

\begin{figure}
	\vspace{-1em}
	\centering
	\subfigure[Cumulative distribution function of total required iterations.]{
		\begin{minipage}[t]{0.8\linewidth}
			\centering
			\includegraphics[width=2.5in]{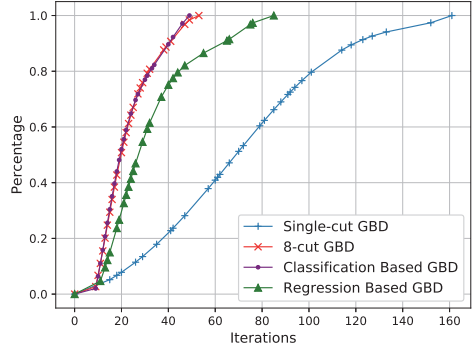}
			\label{fig:gap}
		\end{minipage}
	}
	\subfigure[Cumulative cut number for relaxed master problems.]{
		\begin{minipage}[t]{0.8\linewidth}
			\centering
			\includegraphics[width=2.5in]{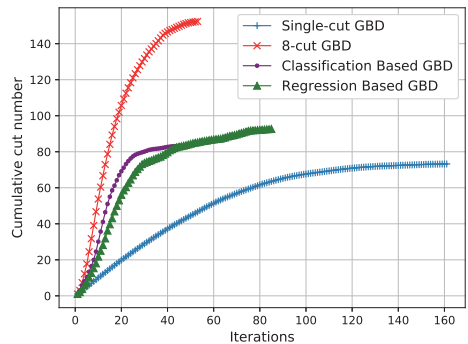}
			\label{fig:cut}
		\end{minipage}%
	}
	\caption{Performance of different GBD algorithms.}
	\label{fig:performance_ML}
\end{figure}

\begin{table}
	\scriptsize
	\caption{Performance of Different GBD Algorithms}
	\vspace{-1em}
	\label{performance}
	\centering	
	\begin{tabular}{|c|c|c|c|c|c|}
		\hline
		Method& \tabincell{c}{Needed \\Iteration \\Number}& \tabincell{c}{Cut \\Number}& \tabincell{c}{Time (s)} &\tabincell{c}{Useful Cut \\Recognition \\Rate}& \tabincell{c}{Useless Cut \\Recognition \\Rate} \\
		\hline
		\tabincell{c}{Single-cut\\ GBD} &73.22 &73.22 &2.97 &/ &/\\
		\hline
		8-cut GBD &23.24&152.37&1.50&/ &/\\
		\hline
		\tabincell{c}{Classification\\ Based GBD} &23.41 &83.49 &0.78 &99.24\% &72.61\%\\
		\hline
		\tabincell{c}{Regression\\ Based GBD} &33.24&92.66&0.98&82.35\% &80.00\%\\
		\hline
	\end{tabular}
\vspace{-2em}
\end{table}

After choosing appropriate techniques for the cut classifier and the cut regressor, respectively, the performance testing results of the proposed method are summarized in Fig. \ref{fig:performance_ML} and Table \ref{performance}. Specifically, Fig. \ref{fig:gap} reflects the cumulative distribution function of the total required iterations. Fig. \ref{fig:cut} shows the cumulative number of the cuts added to the relaxed master problem, which reflects the complexity of the relaxed master problem over the iterations. Table  \ref{performance} gives the detailed values of more performance metrics.
\begin{figure}
	\vspace{-1em}
	\centering
	\includegraphics[width=0.7\linewidth, height=0.2\textheight]{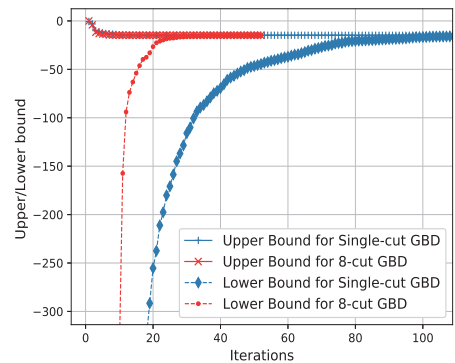}
	\vspace{-1em}
	\caption{Convergence performance of single-cut and 8-cut GBD.}
	\label{fig:convergence}
	\vspace{-2em}
\end{figure}

Before looking into the performance of the ML based GBD, we first pay attention to the comparison between  the single-cut GBD and the 8-cut GBD. The results in Table  \ref{performance} suggest that  the cumulative time for solving relaxed master problems of multi-cut GBD decreases compared with single-cut GBD.  To figure out more details, we depict the convergence performance of the single-cut GBD and the 8-cut GBD in Fig. \ref{fig:convergence}. From this figure, we know that the lower bound of the single-cut GBD, which is related to the relaxed master problems, increases very slowly. It proves that the main bottleneck of the single-cut GBD is the time consumed by solving the relaxed master problems, which accounts for over 90\% of the total computation time \cite{over90}. The multi-cut GBD can reduce the number of required iterations, i.e., the number of relaxed master problems need to be solved, as shown in Fig. \ref{fig:gap}, but it adds more cuts to the relaxed master problems. Hence, the average complexity of each relaxed master problem increases as shown in Fig. \ref{fig:cut}.  Overall, the multi-cut GBD can outperform the single-cut GBD by solving fewer but more complicated relaxed master problems, which suggests that decreasing the number of the relaxed master problems that need to be solved plays a more important role in accelerating GBD than reducing the average complexity of the relaxed master problems. It also indicates that keeping the useful cuts is more helpful than discarding the useless cuts, which is  the reason why we set a higher weight to the class of useful cuts during the training process of the cut classifier. 

As for the classification based GBD, 99.24\% useful cuts are kept and 72.61\% useless cuts are discarded as shown in Table \ref{performance}. Compared with the single-cut GBD, the number of required iterations decreases by 68.03\% while the average complexity of the relaxed master problem only increases by 14.03\%. And the cumulative time for solving master problems decreases by 73.74\%. Compared with the 8-cut GBD, the average complexity of the relaxed master problem decreases by 45.21\% while the number of required iterations only increases by 0.73\%. And the cumulative time for solving master problems decreases by 48.00\%.

As for the regression based GBD, 82.35\% useful cuts are kept and 80.00\% useless cuts are discarded  as shown in Table \ref{performance}. Compared with the single-cut GBD, the number of required iterations decreases by 54.60\% while the average complexity of the relaxed master problem increases by 26.55\%. And the cumulative time for solving master problems decreases by 67.00\%. Compared with the 8-cut GBD, the average complexity of the relaxed master problem decreases by 39.19\% while the number of required  iterations increases by 43.03\%. And the cumulative time for solving master problems decreases by 34.67\%. 

Compared with the accelerated B\&B algorithm in \cite{lorm,learntobranch}, the proposed accelerated GBD algorithm can always achieve the global optimum while the accelerated B\&B  algorithm is a suboptimal one. On the other hand, from the testing results on the scenario with 5 CUs and 3 D2D pairs, our proposed method outperforms the accelerated B\&B  algorithm in terms of practical computational complexity. The accelerated GBD is about twice as fast as the accelerated B\&B algorithm. However, the worst-case complexity of both methods is still exponential due to the NP-harness of MINLP problems.

\subsection{Comparison Between Classification and Regression Based GBD}
In this part, we discuss the advantages and disadvantages of the two proposed ML based GBD. The detailed comparison between these two methods is summarized in Table \ref{comparison}. In terms of methodology, the classification based GBD directly learns the binary labels while the regression based GBD learns the continuous performance indicators. In addition, the data collection method of the classification based GBD is actually a variant of single-cut GBD. However, the  regression based GBD uses multi-cut GBD to collect training data. As mentioned above, the single-cut GBD is more time-consuming than the multi-cut GBD. Therefore, the data collection method of classification based GBD is more time-consuming than that of the regression based GBD. Moreover, both methods can accelerate GBD without loss of optimality. As for the acceleration performance, the classification based GBD performs slightly better than the regression based GBD. There are two possible reasons. On the one hand, the gap between the error rates $\alpha_0$ and $\alpha_1$, i.e., $|\alpha_0-\alpha_1|$, may be very small. As mentioned in Section III.B, the existence of the noise will not influence the classification accuracy if $\alpha_0=\alpha_1$. Therefore, the cut classifier can tolerate the labeling noise and has good performance. On the other hand, the cut classifier has a higher useful cut recognition rate while the cut regressor has a higher useless cut recognition rate. As mentioned above,  keeping the useful cuts is more helpful than discarding the useless cuts. Therefore, the classification based GBD  outperforms the regression based GBD in terms of acceleration performance.
\begin{table}
	\vspace{-1em}
	\scriptsize
	\caption{Comparison Between Classification and Regression Based GBD}
	\vspace{-1em}
	\label{comparison}
	\centering
	\begin{tabular}{|c|c|c|}
		\hline
		Method& Classification Based GBD& Regression Based GBD\\
		\hline
		Methodology& \tabincell{c}{Learning binary label}& \tabincell{c}{Learning continuous \\performance indicator}\\
		\hline
		\tabincell{c}{Data Collection\\ Complexity} & High &Low\\
		\hline
		\tabincell{c}{Useful Cut \\Recognition Ability} &Good &Limited\\
		\hline
		\tabincell{c}{Useless Cut \\Recognition Ability} & Limited &Good\\
		\hline
		Optimality &\tabincell{c}{Globally optimal}&\tabincell{c}{Globally optimal}\\
		\hline
		 \tabincell{c}{Acceleration Performance} &Strong&Good\\
		 \hline
	\end{tabular}
	\vspace{-2em}
\end{table}
\subsection{Generalization Ability of Machine Learning Based GBD}
Generalization ability is a very important property of ML techniques, especially for our proposed method that is expected to be applicable to various mixed integer wireless resource allocation problems. In this part, we test the generalization ability on three different aspects: size aspect, parameter aspect, and problem aspect. Because the cut classifier slightly outperforms the cut regression  in terms of acceleration performance, we only focus on the generalization ability of the cut classifier in the following.

\subsubsection{Generalization Ability on Size Aspect}
First, we test the generalization ability on the size aspect. Specifically, we test how the cut classifier  learned on small-scale scenario performs on the testing problem sets with larger-scale scenarios.  We use the cut classifier learned in Section IV.C, and the testing problem set of each scale includes 50 problem instances. The results are summarized in Table \ref{gene1}. We add three new metrics to evaluate the generalization ability of the proposed method. The first one is the normalized needed number of iterations with respect to that of the multi-cut GBD algorithm. If we can keep all the useful cuts, this metric will be very close to 1. The second one is the normalized cumulative cut number added to the relaxed master problems with respect to that of the single-cut GBD algorithm. If we can discard all the useless cuts, this metric will also be very close to 1. The last one is the speedup in running time with respect  to the multi-cut GBD algorithm, which corresponds to the ultimate goal of the proposed method. Note that the worst-case computational complexity of GBD is exponential, i.e., $O(2^{KL})$, and thus we use $KL$ to characterize the problem's complexity. In Table \ref{gene1}, $KL$ ranges from 15 to 50. The results suggest that the proposed method has a good generalization ability on the size aspect for scenarios with $KL \leq 27$. When we increase the problem size over 27, while the useful cut recognition rate remains excellent,  the useless cut recognition rate drops significantly. Specifically, the useless cut recognition rate drops to 6.43\% for the scenario with 10 CUs and 5 D2D pairs. In this case, the accelerated GBD keeps almost all the cuts and performs similarly to the multi-cut GBD. Therefore, the speedup to multi-cut GBD is quite modest. Overall, the result suggests that the generalization ability on the size aspect is limited to a certain range, which is commonly observed for ML based resource allocation methods  \cite{shi,mlop2,mlop3,lorm,learntobranch,spatiallearning,graphnn,user1}.
\begin{table*}
	\vspace{-3em}
	\scriptsize
	\caption{Generalization Performance on Larger-scale Scenarios}
	\vspace{-1em}
	\label{gene1}
	\centering
	\begin{tabular}{|c|c|c|c|c|c|c|c|}
		\hline
		Testing $(K,L)$& $(5, 3)$& $(9, 2)$& $(6, 4)$&$(7, 3)$ &$(9, 3)$&$(8, 4)$& $(10,5)$\\
		\hline
		\tabincell{c}{Useful Cut Recognition Rate}&99.24\%&99.77\%&99.51\%&99.53\%&99.19\%&99.25\%&99.03\%\\
		\hline
		\tabincell{c}{Useless Cut Recognition Rate} &72.61\%&75.58\%&78.09\%&77.20\%&73.64\%&42.95\%&6.43\%\\
		\hline
		\tabincell{c}{Normalized Needed Iteration Number} &1.00&1.00 &1.07&1.00&1.15&1.05&1.03\\
		\hline
		\tabincell{c}{Normalized Cumulative Cut Number} &1.14&1.10&1.15&1.12&1.22&1.30&1.37\\
		\hline
		\tabincell{c}{Speedup to Multi-cut GBD}&1.92x&1.69x&1.72x&1.73x&1.64x&1.45x&1.12x\\
		\hline
	\end{tabular}
\vspace{-1em}
\end{table*}

\subsubsection{Generalization Ability on Parameter Aspect}
Furthermore, we test the generalization ability on the parameter aspect. We still use the cut classifier in Section IV.C but the D2D network parameters of the testing samples are different from those in Table \ref{para}. Specifically, we change the cell radius of the D2D network,  the maximum transmit power of CU, and the minimum data rate requirement of CU, respectively. The testing problem set of each different parameter setting still includes 50 problem instances and the testing results are summarized in Table \ref{gene2}. From Table \ref{gene2}, it is clear that the proposed method also has a strong generalization ability on the parameter aspect. All the performance metrics remain almost the same for scenarios with different parameter settings. The result suggests that we do not need to train a new model while the network parameters change.
\begin{table}
	\scriptsize
	\caption{Performance with Different D2D Network Parameters}
	\vspace{-1em}
	\label{gene2}
	\centering
	\begin{tabular}{|c|c|c|c|c|}
		\hline
		\tabincell{c}{Testing \\Parameter Setting} &\tabincell{c}{Original\\ Parameters} &\tabincell{c}{Cell Radius\\ as 750 m}& \tabincell{c}{$P^\text{C}_{\max}$ as\\ 22 dBm}& \tabincell{c}{$R^\text{C}_{\min}$ as\\ 1.5 bit/s/Hz}\\
		\hline
		\tabincell{c}{Useful Cut \\Recognition Rate} &99.24\%&99.40\%&99.93\%&98.84\%\\
		\hline
		\tabincell{c}{Useless Cut\\ Recognition Rate} &72.61\%&73.21\%&72.42\%&72.10\%\\
		\hline
		\tabincell{c}{Normalized Needed\\ Iteration Number}&1.00&1.02 &1.00&1.00\\
		\hline
		\tabincell{c}{Normalized Cumulative\\ Cut Number}&1.14&1.15&1.12&1.10\\
		\hline
		\tabincell{c}{Speedup to \\Multi-cut GBD}&1.92x&1.82x&1.69x&1.88x\\
		\hline
	\end{tabular}
\vspace{-1em}
\end{table}

\subsubsection{Generalization Ability on Problem Aspect}
Finally, we test the generalization ability on the problem aspect. Specifically, we aim to test the performance of the proposed method on problems with different objective functions. Note that the testing problems with new objective functions still need to satisfy the convexity and linear separability mentioned in Section II.B. We still use  the cut classifier learned in Section IV.C. The objective function of the training problems is to maximize the minimum data rate among all the D2D pairs as mentioned in Section IV.A. However, the objective functions of the testing problems are different from that. Specifically, we choose two kinds of testing problems. The objective function of the first one is to maximize the sum rate of all the D2D pairs, i.e.,
$$\max_{\{\bm{p^\text{C}}, \bm{p^\text{D}},\bm{\rho}\}} \sum_{l\in \mathscr{L} } R_{l}^\text{D}(\bm{p^\text{C}},\bm{p^\text{D}},\bm{\rho}),$$
and that of the second one is to maximize the weighted sum rate of all the D2D pairs, i.e.,
$$\max_{\{\bm{p^\text{C}}, \bm{p^\text{D}},\bm{\rho}\}} \sum_{l\in \mathscr{L} } \omega_l R_{l}^\text{D}(\bm{p^\text{C}},\bm{p^\text{D}},\bm{\rho}),$$
where $\omega_l$ denotes the weight for the $l$-th D2D pair and is generally selected according to priority in advance. In this test, $\omega_l$ is simply randomly generated according to the uniform distribution between $(0,1)$. 

During the test, we set the number of CUs, $K$, as 5, and that of D2D pairs, $L$, as 3. All other parameter settings are the same as those in Table \ref{para}. We still report the average performance over 50 testing problem instances and the results are summarized in Table \ref{gene3}. The results suggest that the proposed method performs well for scenarios with different objective functions and has a strong generalization ability on the problem aspect. 

Based on the testing results in Tables \ref{gene1}, \ref{gene2}, and \ref{gene3},  we conclude that our proposed method has limited generalization ability on the size aspect but good generalization ability on both the parameter and problem aspects. Therefore, one can start by solving a small convex problem, and then generalize it to other similar convex problems with different objective functions and parameter settings within certain scales.  In this way, one can  avoid generating training samples for new problems, which is part of the overhead mentioned in Section III.B.
\begin{table}
	\scriptsize
	\caption{Performance  with Different Objective Functions}
	\vspace{-1em}
	\label{gene3}
	\centering
	\begin{tabular}{|c|c|c|c|c|}
		\hline
		Objective Function &\tabincell{c}{Max-min \\Problem} &\tabincell{c}{Maximize\\ Sum Rate}& \tabincell{c}{Maximize\\ Weighted\\ Sum Rate}\\
		\hline
		\tabincell{c}{Useful Cut Recognition Rate}  &99.24\%&99.82\% &99.79\%\\
		\hline
		\tabincell{c}{Useless Cut Recognition Rate}&72.61\%&78.07\%& 81.40\%\\
		\hline
		\tabincell{c}{Normalized Needed Iteration Number}&1.00&1.00 &1.03\\
		\hline
		\tabincell{c}{Normalized Cumulative Cut Number}&1.14&1.02& 1.01\\
		\hline
		\tabincell{c}{Speedup to Multi-cut GBD}&1.92x&1.88x& 1.82x\\
		\hline
	\end{tabular}
	\vspace{-2em}
\end{table}

\section{Conclusion}
GBD is a widely used globally optimal algorithm for MINLP resource allocation problems in wireless networks. However, a direct implementation of GBD suffers from high complexity due to the time-consuming master problems. Therefore, this paper incorporates ML techniques to accelerate GBD by decreasing the complexity of the master problems. Specifically, we deal with the acceleration task with two different ML techniques: classification and regression. In this way, a cut classifier and a cut regressor are learned in the supervised manner, respectively. The learned models can distinguish between useful and useless cuts, and only useful cuts are added to the master problem. Therefore, the complexity of the master problem decreases and GBD can be further accelerated. By using a resource allocation problem in D2D communication networks as an example, we verify that the proposed method can reduce the computational complexity of GBD without loss of optimality. Meanwhile, all designs of the proposed method are invariant for different problems, and thus it has good generalization abilities on different aspects, which is a preferred property in wireless networks.

This paper is just a very first attempt to accelerate the solving process of the MINLP resource allocation tasks circumventing problem-dependent designs. Some open problems remain for future work. First, for MINLP problems that do not satisfy the convexity condition, modifications on GBD are needed to guarantee optimality. How to generalize our proposed method to these modified GBD algorithms is an interesting problem worth further investigation. Secondly, our proposed method is implemented in the supervised manner where labeled training samples are needed. However, labeled training samples are usually difficult to obtain in wireless networks. How to decrease the number of needed training samples is an interesting research problem. Moreover,  all designs in our proposed acceleration method are  heuristic, such as the features proposed in Section III.B.  These heuristic designs will not influence the optimality of the accelerated GBD, but there is still room to further speed up the GBD algorithm. Adding some other problem-independent features in the proposed method may lead to better performance. On the other hand, while focusing on a certain resource allocation problem, some problem-dependent features may help reinforce the performance. Using Problem (\ref{Md2d}) as an example, the D2D pair activation ratio, defined as $ \sum_{l\in \mathscr{L} }\rho_l^{(i-1)}/L$, can be used as a problem-dependent feature for the cut generated at the current solution ($\bm{p^\text{C}}^{(i)}$, $\bm{p^\text{D}}^{(i)}$, $\bm{\rho}^{(i-1)}$).  Higher D2D pair activation ratio generally leads to higher overall throughputs. Unfortunately,  there is no general method to design such problem-dependent features.  Usually trial and error is inevitable to find out the key parameters related to the optimization goal in the specific wireless network. Finally, our proposed cut-clean up strategies are still heuristic. One can consider designing other effective methods to further accelerate GBD, and even adopting new ML techniques for wireless MINLP resource allocation problems.

\end{document}